\documentclass{vietnam}

\def\be{\begin{equation}}
\def\ee{\end{equation}}
\def\bea{\begin{eqnarray}}
\def\eea{\end{eqnarray}}

\newcommand{\Photo}{\includegraphics[height=30mm]{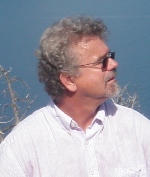}}

\begin{document}
\vspace*{4cm}
\title{Challenges for precision measurements at the LHC$^{\star}$}

\author{ M. W. Krasny}

\address{LPNHE, 
          Universit\'e Pierre et Marie Curie -- Paris 6, Universit\'e Paris Diderot -- Paris 7, \\
          CNRS--IN2P3, 4 pl. Jussieu, 75005 Paris, France} 
 
\vspace{0.5cm}          

\maketitle\abstracts{
Challenges for precision measurements at the LHC are discussed and a proposal how to 
move forward to  overcome the LHC-specific precision brick-walls is  presented.}

\footnoterule
\noindent
{\footnotesize
$^{\star}$The work presented in this contribution has been  partly supported by the program of the French--Polish 
co-operation between IN2P3 and COPIN no.\ 05-116 and \ 05-117.}

\section{What I mean by precision measurements}

In the presented contribution  the ``precision measurements" are those which test the Standard Model (SM) 
with a better precision with respect to what  has so far been achieved.
We shall not discuss here the measurements for which the statistical errors are the dominant ones 
(e.g. the measurements of  rare decay modes of the of heavy flavours),  but focus our attention 
only  on those of the  measurements for which the  precision is limited by the systematic measurement errors  
or by the Monte-Carlo modeling uncertainties. The latter may include both the uncertainties in the 
modeling of the parton distribution functions (PDFs) and the uncertainties inherent to  the theoretical framework, 
used in the unfolding procedures (the order of the perturbative expansion, the presence, or lack, of the 
higher twist  (HT) effects, etc.). 
  
Up to now  the LHC has not contributed significantly to the domain of precision measurements.   
The questions we shall try to answer  in this contribution are: (1)  Why?  and 
(2) What could be the way forward for the LHC precision measurement programme 
to be competitive?

\section{Challenges for precision physics at the LHC}

The hadronic colliders are optimal for generic exploration of interactions of a large variety of
the Standard Model point-like particles, over a large momentum range, which extends up to the
momentum of the colliding hadrons. Their merits are complementary to those of
electron--positron colliders, which employ better controlled, but less luminous  beams of the point-like 
particles, colliding in a cleaner environment.

It is obvious to everyone that the  hadronic colliders can hardly compete with the leptonic ones 
in the measurement precision for those of the  SM processes for which the systematic errors dominate over the statistical ones. 
What, however, often  escapes attention, and is a focus point of  this contribution, is that the LHC proton--proton collision scheme 
is by far more challenging than  the proton--antiproton collision one.  

At the LHC the symmetry relations (specific to the proton--antiproton colliders) are no longer at work. We thus 
need to  understand the charge and polarisation asymmetries in the $W$ and $Z$ boson  production processes to a much higher precision 
than at the Tevatron. As a consequence,  the relative strength  of the valence and sea contributions to the proton wave function 
must be understood to a much higher precision at the LHC 
than at the Tevatron.  

The LHC beams are accelerated to a much  higher energy
than at the Tevatron ones. The heavy flavour excitations of the proton become thus significant and  
we shall have to understand the heavy flavour content of the proton wave function to a  much better precision
than that required for the Tevatron measurements.  

The  $W$ and $Z$ bosons are produced at the LHC  by the low-$x$ partons. We thus 
need to precisely understand not only the $x$ dependence of their distributions and  their momentum, 
flavour and spin correlations but,  in addition,   their  flavour-dependent transverse momentum  distributions. 

Last but not the least,  the gain in the collected luminosity  at the LHC with respect to the Tevatron is achieved at the cost of 
a  large $pp$ collision pile-up. 

In view of all the LHC-specific problems the following question may be asked:
Can the LHC experiments improve the measurement precision 
achieved at the previous colliders?  

The departure point of this contribution is the following statement: 
{\bf If  the  Monte Carlo tools and the measurement procedures developed  for  the Tevatron programme are 
used at the LHC, and if no LHC-specific effort is  undertaken, then the precision of 
a large majority of the  SM measurements will not be improved at the LHC -- no matter what level 
of understanding of the LHC machine and detectors performance will eventually be achieved}. 

The goal of this contribution  is to introduce   several ideas how to move forward to overcome these LHC-specific measurement precision brick-walls. 
These ideas were developed and quantitatively evaluated in  the series of  papers which are recalled in the next section of this contribution. 
  
\section{How to move forward}

\subsection{ Use of flexibilities of operation modes of the LHC}

The following operation modes of the LHC, most of them not feasible  at the Tevatron, 
could be used to significantly improve the precision of the LHC measurements:

\begin{itemize}
\item 
running at several, suitably chosen, centre-of-mass energies  to reduce the impact of systematic experimental  errors
(e.g. the jet energy scale), and the modeling errors (e.g. missing higher order QCD corrections) on the measured quantities -- 
proposed and discussed in  ref.~\cite{CM1}, ref.~\cite{CM2} and more recently in  ref.~\cite{CM3};
\item 
running, for a fraction of the LHC operation time,  the proton--ion collisions for a precision investigation of the 
electroweak vacuum properties -- proposed and discussed in ref.~\cite{pA};
\item 
running, for a fraction of the LHC operation time,  iso-scalar ion beams, such as the deuterium or helium ones, to restore the strong isospin symmetry 
of the light valence and sea quarks and, as a consequence,  to get rid of he corresponding PDF modeling uncertainties -- proposed in ref.~\cite{ISO};
\item 
running, for a small fraction of time, partially stripped ion beams at the LHC to deliver the monochromatic electron beam to the interaction points of the LHC experiments 
(e.g. for precision calibration of the detector response to jets)  -- see ref.~\cite{eBEAM} for the feasibility studies of such a running mode; 
\item 
using, for the experimental control of the pile-up effects, the ``precision oriented" LHC bunch-filling scheme in which the first two Ò3-batchÓ bunch trains,  injected from the SPS to the LHC,  have a reduced number of protons to assure, on the average, one collision per bunch crossing (the  remaining bunches injected in the same LHC fill are stored  at their nominal proton density)  -- such a LHC filling scheme was proposed in ref.~\cite{Myers}. 
\end{itemize}

\subsection{Creation of  Òprecision supportÓ LHC-auxiliary exp. programme}

There are several measurements (e.g. the precision measurement of the $W$-boson mass) where the the ultimate precision 
will be limited by the accuracy of the LHC-external input which is necessary to derive 
the values of the SM parameters  from the LHC measurements. 
Two initiatives can be mentioned here:
(1) the letter of intent for the dedicated SPS fixed target experiment~\cite{SPS},  having as a main goal 
a high-precision understanding of the $W$ and $Z$ polarisation at the LHC, and (2)  the initial proposal of 
the iCHEEPx  project~\cite{iCHEEP} to collide electrons coming from the 
2.45 GeV Energy Recovery Linac,  with 6 recirculation passes in the arcs, providing the electron beams of: 5.5, 10.4, 15.3, 20.2, 25.1 and 30.0 GeV 
with the SPS proton and ion beams (over their full momentum range). The principal  target of the latter initiative 
is to understand the QCD processes at the femto-meter distance scale.

\subsection{Switching  to a new operation mode of the LHC experiments}

The present configuration of the trigger, online event selection, and event reconstruction framework, which 
the LHC experiments inherited from the previous large scale HEP experiments,  may  turn out to be inadequate 
for the precision measurement phase of the LHC. The present  configuration  gives too a little freedom for  the physics groups 
to implement their, specific task oriented, data selection, reconstruction  and analysis methods.

The Gauge Model of the Trigger, Data Acquisition and the Data Analysis for the LHC experiments proposed in ref.~\cite{GAUGE}
defines  an alternative configuration. In such a configuration  physics groups may implement  their own on-line and off-line data handling schemes which  
may run concurrently in a mutually transparent way. 

\subsection{Precision oriented upgrade the LHC detectors }

One of the most important upgrade task for the LHC experiments to increase the accuracy 
of their measurements is to construct and implement a dedicated detector capable to cross-normalize:
\begin{itemize}
\item
the cross sections at different centre-of-mass energies, 
\item  
the cross sections measured in runs with different beam species, 
\item 
the cross sections measured at the LHC and at the Tevatron,
\end{itemize}
with a LEP-like per-mille precision. 
A per-mille cross-normalisation of the cross-sections is of primordial importance for e.g. the filtering out the  Primakoff processes, Higgs sector and 
Quark-Gluon-Plasma measurements. The  detector design and the corresponding measurement strategy capable to achieve 
the requisite precision has been proposed and evaluated in  ref.~\cite{LUMI1}, ref.~\cite{LUMI2} and ref.~\cite{LUMI3}.

\subsection{Define new observables and new precision measurement strategies}

The observables measured at the LHC are sensitive not only to the underlying physics 
mechanisms,  but also to the detector systematic effects and to the approximations present in the models 
which are used both to  interpret the measurements  and to unfold the ``truth" distributions. 
The goal of introducing new observables and new measured strategies is to try maximize their sensitivity to the 
physics effects, while minimizing the effects of the systematic error sources.
The concrete examples of the new measurement strategies and of the new observables 
scan be found in ref.~\cite{CM1}, ref.~\cite{CM2}, ref.~\cite{OBS} and ref.~\cite{ISO}. They  
exploit the flexibility of the detector and the machine running modes to  minimize the impact 
of the dominant sources of the systematic and modeling errors
on the precision of the measurement of the SM parameters such as: $M_W$, $\Gamma _W$, their 
charge asymmetries, $\sin \theta _W$, and $\alpha _s$.

\section{Conlusions}

The measurements at the LHC require already a substantial effort to reach the precision limits of the LHC detectors and to reach the requisite  precision of the event generators. The basic message of  this contribution is that such an initial  effort is necessary but, unfortunately,  not sufficient.
In order to be competitive, new measurement strategies exploiting fully the flexibilities of the LHC collider and its detectors (e.g. such as sketched in this talk) must be incorporated. The precision measurement programme is worth an effort, not only for the textbook measurements,  but also as a complementary approach 
to searches for new physics phenomena at the precision frontier of particle physics.     

\section*{Acknowledgements}

The studies serving as a base for this contribution were made in collaboration with W. P\l{}aczek, F. Fayette, A. Si\`odmok, F. Dydak, S. Jadach, J. Chwastowski and 
K. S\l{}owikowski.

\end{document}